# Cybersecurity Cost of Quality: Managing the Costs of Cybersecurity Risk Management

Nicole Radziwill and Morgan Benton

## Abstract

There is no standard yet for measuring and controlling the costs associated with implementing cybersecurity programs. To advance research and practice towards this end, we develop a mapping using the well-known concept of quality costs and the Framework Core within the Cybersecurity Framework produced by the National Institute of Standards and Technology (NIST) in response to the Cybersecurity Enhancement Act of 2014. This mapping can be easily adopted by organizations that are already using the NIST CSF for cybersecurity risk management to plan, manage, and continually improve cybersecurity operations. If an organization is not using the NIST CSF, this mapping may still be useful for linking elements in accounting systems that are associated with cybersecurity operations and risk management to a quality cost model.

## Introduction

In 1995, The American Society for Quality (ASQ) presented an article in its flagship publication *Quality Progress* about "cyberquality" - defined as *information about quality you can find on the internet*. Because technology has progressed by orders of magnitude over the past two decades, this use of the term now seems overly simplistic! There is a new "cyberquality" which is readily apparent in the ISO 9001 definition of quality: "the totality of characteristics of an entity that bear upon its ability to satisfy stated and implied needs." If that entity is a networked, connected device or system, then the stated and implied needs of customers and other stakeholders cannot be met without cybersecurity.

With the growth and evolution of the Internet of Things (IoT), people, vehicles, homes, city infrastructures, and industrial infrastructures are becoming more tightly interconnected, requiring new strategies for designing quality into systems. (Radziwill & Benton, 2017) At the same time, we will need more robust metrics for controlling and continuously reducing the costs associated with both quality assurance and cybersecurity risk management. Cyberquality will become the net effect of simultaneously meeting quality goals and cybersecurity goals, so it makes sense to explore cost metrics that are linked to both domains.

As the number of networked components increases, so does the potential for catastrophic impact. Over the last several years, reports of software deployed over the Internet designed to destroy infrastructure have increased - and infrastructure impacts all organizations. Such "cyber attacks" typically occur when software is deployed, usually (but not always) via a network, onto other systems - with the intent of destroying physical or information assets. These attacks occur via a number of attack vectors, and not all are technological: systems can also be breached maliciously or accidentally by insiders, and "social engineering" can be used to trick trusted users into giving up their passwords or answers to security questions. With nearly 26 million IoT endpoints expected by the year 2019, the pace and intensity of intrusions will increase. (Trautman & Ormerod, 2017)

Because "repeated cyber intrusions into critical infrastructure [demonstrated] the need," in 2013, President Obama issued Executive Order 13636 (EO 13636) directing the National Institute for Standards and Technology (NIST) to create a "framework to reduce cyber risks to critical infrastructure" (Obama, 2013, p. 11739). In the context of this order, "critical infrastructure" refers to:

> *"... systems and assets, whether physical or virtual, so vital to the United States that the incapacity or destruction of such systems and assets would have a debilitating impact on security, national economic security, national public health or safety, or any combination of those matters."* (Obama, 2013, p. 11739)

This paper explores the intersection of the NIST Cybersecurity Framework (CSF) and quality cost models, with the goal of making it easier for organizations to study, monitor, and control the costs associated with cybersecurity. Although the NIST CSF was intended to support the critical infrastructure systems that most production systems rely on, it is broadly applicable for cybersecurity risk management at organizations of all sizes.

## Background

There are four topics that inform the mapping that serves as the primary contribution of this article. They are: cybersecurity economics, the NIST CSF, the concept of quality costs, and models that describe how quality costs are typically distributed in organizations. These topics are covered in order in the following sections.

**Cybersecurity Economics**

Protecting the confidentiality, integrity, and accessibility of information takes time, effort, and money. Research on the costs of cybersecurity date back nearly two decades, mostly focused on two themes: budgeting appropriately, and determining the economic impacts of cyber attacks. For example, Campbell et al. (2003) examined the economic implications of cybersecurity breaches using stock market performance as an indicator. By creating models that estimated the stock valuation in the absence of an attack, and comparing them to stock performance after attacks, they found detectable dips in stock price only following attacks that involved unauthorized access to confidential data. Attacks that did not appear to have direct impact on the customer were not associated with this same pattern.

Böhme (2010) performed a comprehensive review of the literature to examine relationships between the costs of information security and benefits realized from making those expenditures. This author recognized that there is a baseline level of security provided by preventive efforts for risk mitigation, along with testing those elements. At some point, the costs level out, so that to provide robust mitigation of external breaches, it would cost much more than many organizations are willing to invest. There is "art and science" associated with identifying the ideal balance, so the author recommends using *Return on Security Investment (ROSI)*, which is the benefits less the costs, divided by the costs, and converted into a percentage.

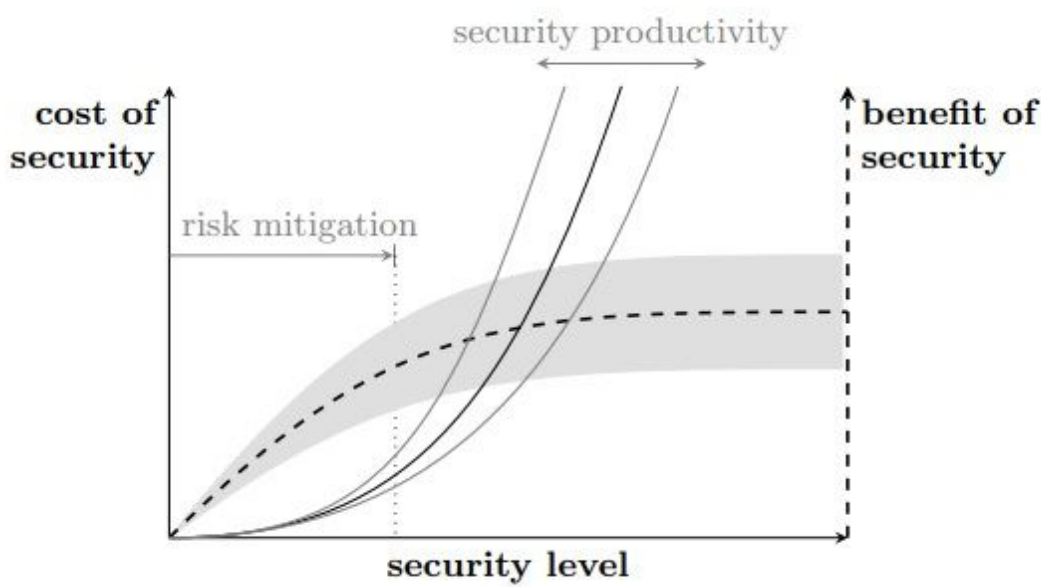

**Figure 1.** Cost/benefit relationships in information security. From Böhme (2010).

Brecht & Nowey (2013) reviewed all techniques that had been applied to assessing costs of information security, and categorized them into four areas: cost/benefit analysis of cybersecurity (including research on optimal investment), cost of cybercrime, surveys summarizing the actual costs of cybersecurity management, and quality cost models. The category on quality cost models was the only one that did not cite research directly related to the domain of information security, but rather seemed to suggest that applying quality cost models would be a logical next step. They suggest that effective cost models in cybersecurity will address the costs of purchasing, operating, implementing, and depreciating tools and systems; costs of operating those tools and systems; costs of consulting or other labor; and cost of risk or uncertainty.

Gordon et al. (2011) examined the impact of cyber attacks on stock returns, and found that when grouped according to the three tenets of information security (confidentiality, integrity, and availability), breaches that impacted availability have the greatest negative effects. Gordon et al. (2015) extended the Gordon-Loeb (GL) Model for determining the optimal level of investment in cyber security activities to account for externalities such as botnets (global networks of infected computers that can be used to launch denial of service attacks, or compromised individuals that engage in psychological warfare on behalf of a nation-state, terrorist, or activist group). With externalities considered, they found that most organizations underinvest in cybersecurity operations, and affirm that "governments around the world are justified in considering regulations and/or incentives designed to increase cyber security investments by private sector firms."

Moore et al. (2016) interviewed 40 executives with primary responsibility for cybersecurity, selected from Chief Information Security Officers (CISOs) and Chief Information Officers (CIOs) drawn mainly from healthcare, financial, retail, and government. 31 of the respondents were from the US, and 9 were international. The questions, which were exploratory in nature, focused on how threats were identified, prioritized, and managed, and the decision-making process for cybersecurity investments at the respondents' organizations. They noticed few differences between the industry sectors, and remarked that finding qualified cybersecurity professionals seemed, in general, to be much more challenging that finding funding to support cybersecurity.

These authors also drew a conclusion that is directly relevant to the need for the present study. They report that in practice, "there is much less focus on the actual results of cybersecurity efforts, such as examining costs and the effectiveness of controls. This may be due to the widespread use of frameworks [such as the NIST Cybersecurity Framework] which promote the use of process measures." A mechanism for examining costs that is aligned with a process-based framework thus may have rather broad applicability.

**The NIST Cybersecurity Framework (CSF)**
The Cybersecurity Enhancement Act of 2014 further clarified the intent of EO 13636, directing the National Institute of Standards and Technology (NIST) to "facilitate and support the development of a voluntary, consensus-based, industry-led set of standards, guidelines, best practices, methodologies, procedures, and processes to cost-effectively reduce cyber risks to critical infrastructure." (S.1353, 2014) Responding to this charge, NIST published three requests for information (RFI) to learn how organizations were managing cybersecurity risk, and to identify best practices. NIST published its first version of the CSF, which captured and organized the results, in February 2014.

The NIST CSF provides proactive risk-based guidance and a common, technology-neutral language for cybersecurity management. It complements an organization's cybersecurity operations, can be used to launch a cybersecurity program where none exists, and can be used in conjunction with other standards and guidance, including ISO 31000 (Risk Management), the ISO/IEC 27000 series (Information Security Management Systems), and NIST Special Publication (SP) 800-39 (Managing Information Security Risk). Although designed with the protection of critical infrastructure (power generation, water/wastewater management, transportation systems) in mind, it can be applied to manage cybersecurity risk in any environment. The framework is voluntary, not prescriptive, and can be used with many different risk management tools, techniques, and practices.

NIST CSF is a toolkit of 98 "pointers" to guidance provided by five standards or collections of best practices. The pointers, called "subcategories," are classified into five functions: **Identify**, **Protect**, **Detect**, **Respond**, and **Recover**. The five functions are further broken down by tasks in cybersecurity risk management, which are referred to as categories. The table of functions, categories, subcategories, and informative references is called the Framework Core.

Figure 2 shows how the Framework Core begins. The first category, "Asset Management," contains 6 of the 98 total subcategories ("pointers" to the standards and guidance in the far right column). Each subcategory represents an objective, task, or group of tasks that must be performed to advance cybersecurity risk management. For example, ID.AM-1 is "Physical devices and systems within the organization are inventoried." This specifies *what* must be done, not *how* it should be done. The "Informative References" column on the far provide directs the NIST CSF user to applicable areas of standards and guidance that could be considered as the organization is deciding *how* to inventory devices and systems.

| Category | Subcategory | Informative References |
|---|---|---|
| **Asset Management (ID.AM):** The data, personnel, devices, systems, and facilities that enable the organization to achieve business purposes are identified and managed consistent with their relative importance to business objectives and the organization's risk strategy. | **ID.AM-1:** Physical devices and systems within the organization are inventoried | · CCS CSC 1<br>· COBIT 5 BAI09.01, BAI09.02<br>· ISA 62443-2-1:2009 4.2.3.4<br>· ISA 62443-3-3:2013 SR 7.8<br>· ISO/IEC 27001:2013 A.8.1.1, A.8.1.2<br>· NIST SP 800-53 Rev. 4 CM-8 |
| | **ID.AM-2:** Software platforms and applications within the organization are inventoried | · CCS CSC 2<br>· COBIT 5 BAI09.01, BAI09.02, BAI09.05<br>· ISA 62443-2-1:2009 4.2.3.4<br>· ISA 62443-3-3:2013 SR 7.8<br>· ISO/IEC 27001:2013 A.8.1.1, A.8.1.2<br>· NIST SP 800-53 Rev. 4 CM-8 |
| | **ID.AM-3:** Organizational communication and data flows are mapped | · CCS CSC 1<br>· COBIT 5 DSS05.02<br>· ISA 62443-2-1:2009 4.2.3.4<br>· ISO/IEC 27001:2013 A.13.2.1<br>· NIST SP 800-53 Rev. 4 AC-4, CA-3, CA-9, PL-8 |
| | **ID.AM-4:** External information systems are catalogued | · COBIT 5 APO02.02<br>· ISO/IEC 27001:2013 A.11.2.6<br>· NIST SP 800-53 Rev. 4 AC-20, SA-9 |
| | **ID.AM-5:** Resources (e.g., hardware, devices, data, and software) are prioritized based on their classification, criticality, and business value | · COBIT 5 APO03.03, APO03.04, BAI09.02<br>· ISA 62443-2-1:2009 4.2.3.6<br>· ISO/IEC 27001:2013 A.8.2.1<br>· NIST SP 800-53 Rev. 4 CP-2, RA-2, SA-14 |
| | **ID.AM-6:** Cybersecurity roles and responsibilities for the entire workforce and third-party stakeholders (e.g., suppliers, customers, partners) are established | · COBIT 5 APO01.02, DSS06.03<br>· ISA 62443-2-1:2009 4.3.2.3.3<br>· ISO/IEC 27001:2013 A.6.1.1<br>· NIST SP 800-53 Rev. 4 CP-2, PS-7, PM-11 |

**Figure 2.** The first category (Asset Management) in the Framework Core. (NIST, 2014b)

There are five standards/guidance that are leveraged by the NIST CSF. Together, they provide a comprehensive platform for unified cybersecurity operations, risk management, and strategic management. These are:

- **Center for Cybersecurity Top 20 Critical Security Controls (CCS CSC)** - a standard for computer security that outlines 20 key actions that can be taken to block or mitigate the effects of known attacks. (CIS, 2016) The 20 controls have been mapped to the five NIST CSF functions. (SANS, 2016)
- **COBIT 5** (ISACA, 2012) - a business framework for management and governance of information security. Used to translate business and customer-focused needs into actionable operations objectives.
- **ISA/IEC 62443-2-1:2009/ISA 62443-3-3:2013** (ISA/IEC, 2013) - standards specifically for the security of industrial control systems, formerly known as ISA99.
- **ISO/IEC 27001:2013** (ISO/IEC, 2016) - standards describing best practices for the management of information security, and also providing a path towards compliance with HIPAA, Sarbanes-Oxley, and Payment Card Industry (PCI) regulations.
- **NIST Special Publication (SP) 800-53 Rev. 4** (NIST, 2013) - security controls and assessment procedures organized into 18 groups, each with its own specific function (e.g. access control, contingency planning, incident response)

An alternative view of the Framework Core is also available. (NIST, 2014c) Shown in Figure 3, this mapping provides the same information as in the Framework Core, but no descriptions are provided and there is a separate column for each reference. This supplement is useful for organizations that already have institutional capabilities aligned with one or more of the references, and do not plan to consult every one. The coverage of subcategories is also more clear in the alternative mapping of the Framework Core than it is in NIST (2014b).

| Function | Category | Subcategory | CCS CSC | COBIT 5 | ISA 62443-2-1:2009 | ISA 62443-3-3:2013 | ISO/IEC 27001:2013 | NIST SP 800-53 Rev. 4 |
|---|---|---|---|---|---|---|---|---|
| ID | AM | AM-1 | CSC 1 | BAI09.01, BAI09.02 | 4.2.3.4 | SR 7.8 | A.8.1.1, A.8.1.2 | CM-8 |
| ID | AM | AM-2 | CSC 2 | BAI09.01, BAI09.02, BAI09.05 | 4.2.3.4 | SR 7.8 | A.8.1.1, A.8.1.2 | CM-8 |
| ID | AM | AM-3 | CSC 1 | DSS05.02 | 4.2.3.4 | | A.13.2.1 | AC-4, CA-3, CA-9, PL-8 |
| ID | AM | AM-4 | | APO02.02 | | | A.11.2.6 | AC-20, SA-9 |
| ID | AM | AM-5 | | APO03.03, APO03.04, BAI09.02 | 4.2.3.6 | | A.8.2.1 | CP-2, RA-2, SA-14 |
| ID | AM | AM-6 | | APO01.02, DSS06.03 | 4.3.2.3.3 | | A.6.1.1 | CP-2, PS-7, PM-11 |

**Figure 3.** Alternative mapping of the Framework Core. (NIST, 2014c)

The NIST CSF links strategic planning, quality management, risk management (e.g. using ISO 31000), and cybersecurity operations. Because it fills the gap between cybersecurity operations and the quality/risk planning efforts that are usually only done at the executive and business process levels, it plays a central role in other tools and frameworks. This includes the Baldrige

Cybersecurity Excellence Builder (BCEB), and the widely-applied Cybersecurity Capability Maturity Model (C2M2) family for domain-specific assessment and continuous improvement of cybersecurity risk management. C2M2 guidance covers electric power generation and distribution, oil/natural gas, and dams. (Miron & Muita, 2014)

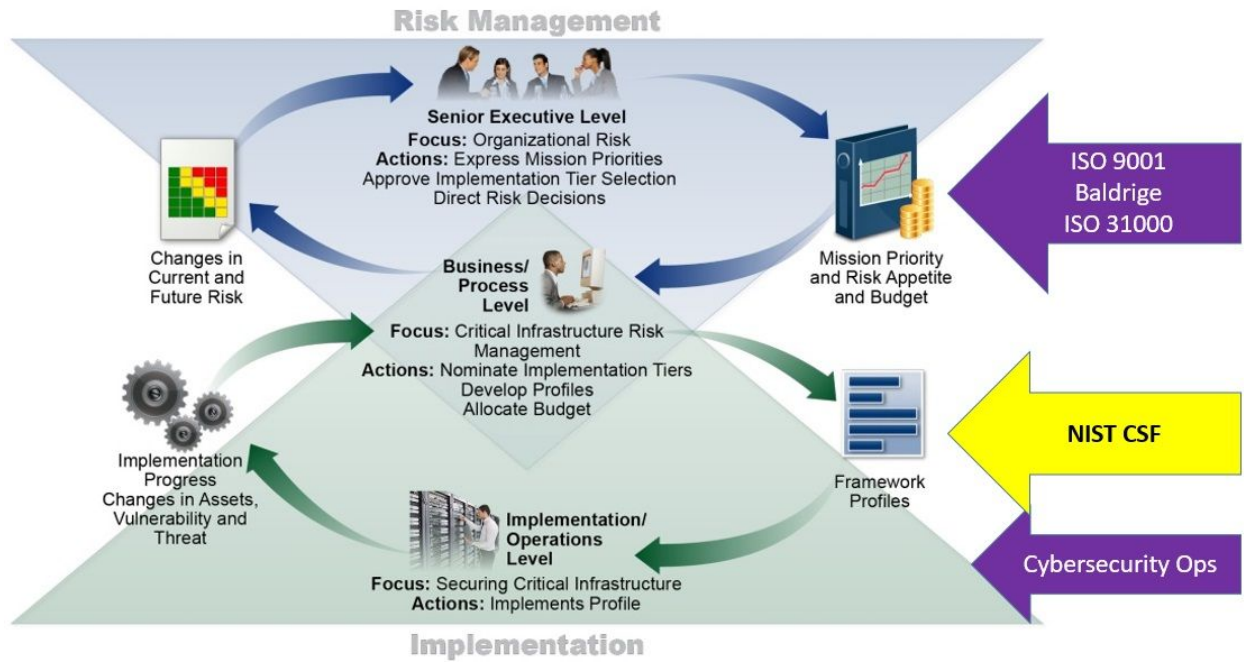

**Figure 4.** NIST CSF complements cybersecurity operations, risk management, and quality management. [Adapted from Fig. 2 in NIST (2014).]

**Quality Cost Models**
Quality is often promoted as a key element for achieving and maintaining competitiveness, and cost of quality metrics can be used to facilitate quality improvements that translate into cost reductions. (Campanella, 1999a) This concept has been applied to manufacturing tangible products, software-intensive products and systems, and components in the Internet of Things (IoT), and can be applied in both development and operations contexts. (Radziwill, 2006)

There are many variations on quality cost models (sometimes referred to as "cost of poor quality") in the literature. All models address the cost of conforming to requirements, and the cost of failing to conform to those requirements; some even include opportunity costs (the costs associated with not taking a certain action). (Schiffauerova & Thomson, 2006) The most commonly used models establish that the *cost of conformance* is the sum of the cost to *prevent* issues and the cost to *test* them (appraisal); the *cost of nonconformance*, also sometimes called *cost of rework*, is the *cost of internal failures* added to the *cost of external failures* - those problems that are recognized, or directly experienced by, customers and other stakeholders. Internal and external failures are distinguished based on *who* is impacted. In many papers, these are referred to as Prevention-Appraisal-Failure (PAF) models. To summarize:

Cost of Quality (CoQ) = Cost of Conformance + Cost of Nonconformance

Cost of Conformance = Cost of Prevention + Cost of Appraisal

Cost of Nonconformance = Cost of Internal Failures + Cost of External Failures

Cost of Quality = Cost of Prevention + Cost of Appraisal + Cost of Internal Failures + Cost of External Failures

Thomas (2009) proposed a "cost of security" model based on the Loss Distribution Approach (LDA) that mentions costs of quality, but does not build on its models. Brecht & Nowey (2013) and Böhme (2010), in their studies of costs in information security, both identified that quality cost models may be appropriate for cybersecurity. Böhme (2010), however, did not mention quality costs directly, but identified the quality cost categories in his articulation of cost/benefit relationships in information security. In Figure 5, "protection measures" are prevention, "qualitative evaluation" and "penetration testing" are appraisal, "incident counts" could be categorized by whether the incidents had internal or external impact, and "direct loss" refers to external failures.

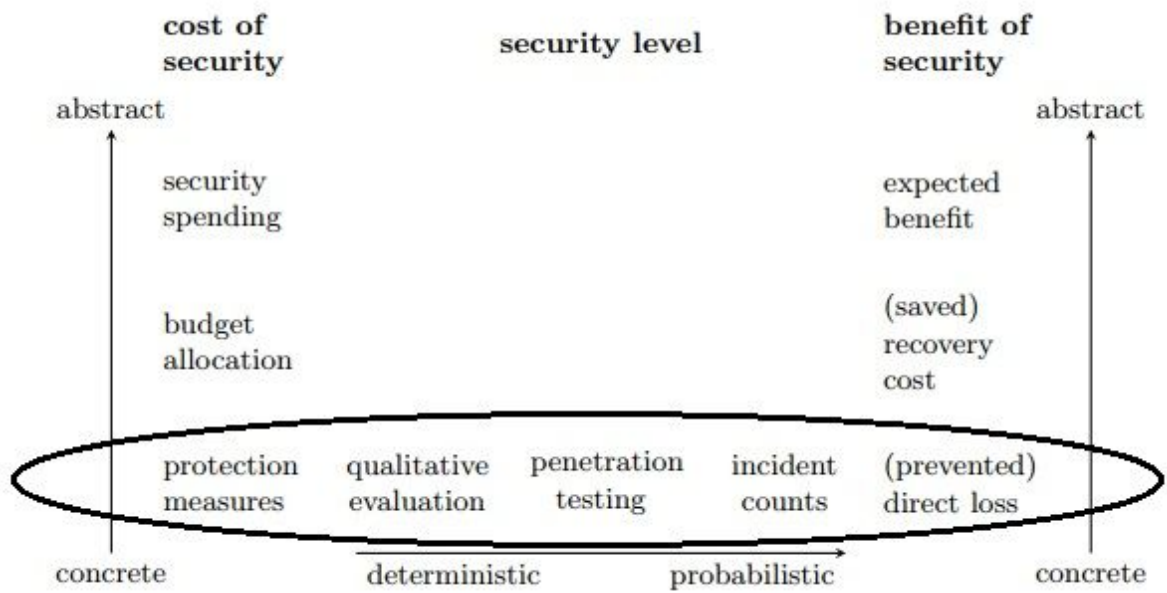

**Figure 5.** Quality costs within cost/benefit relationships in information security. Adapted from Böhme (2010).

Although similar to the production of tangible products in many ways, the development of software is an economically unique activity. Table 1 identifies some of the differences.

Cybersecurity shares many of the same characteristics as software-intensive production: in particular, specifications change extremely rapidly, and defects are related to human understanding of the current state of threats, vulnerabilities, and capabilities.

| Product Manufacturing | Software-Intensive Production |
| --- | --- |
| Physical product | Intellectual product |
| Output is subject to physical laws and constraints | Output is subject to human constraints and logical constraints |
| Specification is stable | Specification is constantly changing |
| Product defects more often the result of faulty materials, machines, or inspection | Product defects more often the result of human mistakes and misunderstandings, or not anticipating ways in which product will be used |
| Effectively executing processes to satisfy requirements is key | Understanding requirements is key |
| Marginal cost associated with producing more units of a product | No marginal cost for producing additional product |

**Table 1.** Differences between product manufacturing and software-intensive production.

Because of these similarities, and also because so many security controls revolve around software or the software development lifecycle (e.g. CCS CSC 2) the PAF model for quality costs can be extended to include development activities. In the context of cybersecurity operations, the assumption is made that any of the items that contribute to cost of quality also contribute to cost of cybersecurity. As a result, an executive-level dashboard might show two values for cost of quality: one specific to activities that enhance cybersecurity, and one for other quality-related costs that are not specific to cybersecurity.

Figure 6 shows how costs of quality are related to total development costs. If an organization measures cybersecurity cost of quality (CCoQ) in addition to ordinary cost of quality, there will be an additional branch off total development costs. Even though there is only one block in Figure 6 that expressly calls out cost of labor, it is likely that many of the blocks will be dominated by labor costs because of the unique economic aspects of developing software-intensive systems discussed earlier.

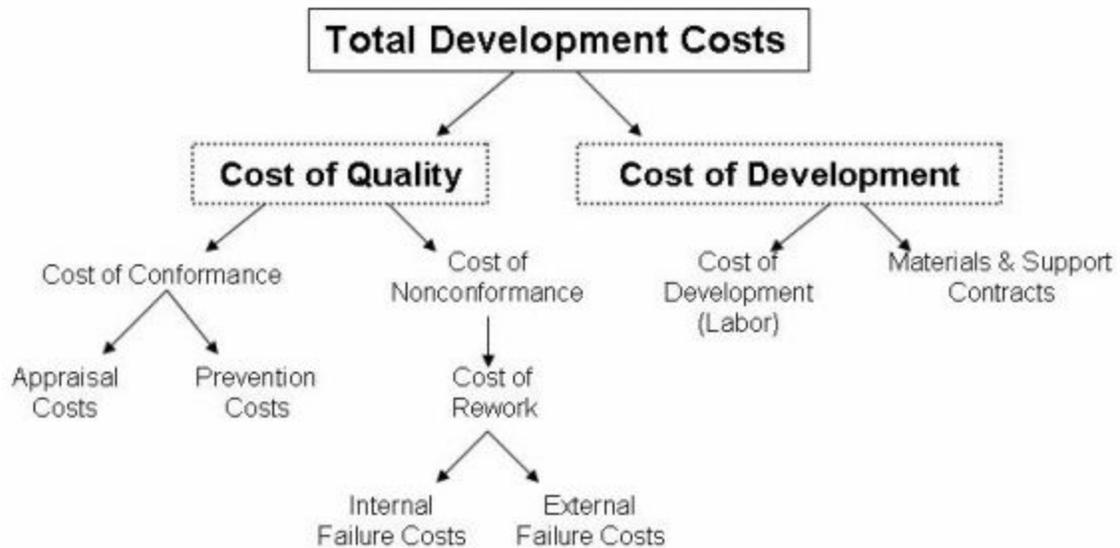

**Figure 6.** Quality costs applied to software development. (From Radziwill, 2006)

**Quality Costs in Practice**
This model provides great flexibility for organizations that want to examine opportunities for improving resource utilization from several perspectives at once:

Total Development Costs = Cybersecurity Cost of Quality (CCoQ) +
Cost of Quality (CoQ) + Cost of Development

There are many different ways to use this data. Total development costs, or any of its three components, can be tracked longitudinally on a monthly or quarterly basis. Only cybersecurity cost of quality can be tracked, and this can still add value. Total cost of quality, ordinary cost of quality, and/or cybersecurity cost of quality can be expressed as a percentage of total development costs and tracked over time.

For best results, however, data should be collected at the lowest levels of the hierarchy (appraisal, prevention, internal failures, and external failures in cost of quality; cost of labor, cost of materials and contracts in cost of development). With data organized this way, any of the intermediary categories can also be tracked (e.g. cost of nonconformance vs. cost of conformance, or cost of rework). Important patterns that will help managers facilitate improvements to cybersecurity programs can be found in any of these elements, and so some experimentation will be required.

Examples of cybersecurity-related activities that fall into each of the quality cost categories is shown in Table 2. The example activities were drawn from the Center for Internet Security's Top

20 Critical Security Controls, which provides guidance in the NIST CSF. This list is meant to be illustrative, not exhaustive.

| Quality Cost Category | Cybersecurity Activities (Selected from CIS, 2013) |
|---|---|
| **Prevention** | CSC 1: Inventory of authorized and unauthorized devices<br>CSC 2: Inventory of authorized and unauthorized software<br>CSC 3: Secure configurations for Hardware and software<br>CSC 5: Control use of administrative privileges<br>CSC 7: Email & web browser protections<br>CSC 8: Malware defenses<br>CSC 9: Limit & control ports, protocols, services<br>CSC 11: Secure configurations for network devices<br>CSC 12: Boundary defense<br>CSC 13: Data protection<br>CSC 14: Controlled access based on need-to-know<br>CSC 15: Wireless access control<br>CSC 16: Account control & monitoring<br>CSC 18: Application software security |
| **Appraisal** | CSC 3: Test secure configurations for hardware and software<br>CSC 4: Continuous vulnerability assessment & remediation<br>CSC 5: Test use of administrative privileges<br>CSC 6: Maintenance & monitoring of audit logs<br>CSC 7: Test email & web browser protections<br>CSC 8: Test malware defenses<br>CSC 9: Test limiting of ports, protocols, services<br>CSC 10: Test data recovery<br>CSC 11: Test secure configurations for network devices<br>CSC 12: Test boundary defense<br>CSC 13: Test data protection<br>CSC 15: Test wireless access control<br>CSC 17: Security skills assessment<br>CSC 18: Test application software security<br>CSC 20: Penetration tests & red team exercises |
| **Internal Failures** | CSC 4: Continuous vulnerability assessment & remediation - respond to internal breaches and notifications from agencies that monitor vulnerabilities<br>CSC 5: Respond to internal misuse of administrative privileges<br>CSC 10: Respond to internal data recovery issues<br>CSC 12: Respond to internal boundary defense issues<br>CSC 15: Respond to internal issues that result from wireless access control<br>CSC 19: Internal incident response |
| **External Failures** | CSC 4: Continuous vulnerability assessment & remediation - respond to external breaches<br>CSC 10: Respond to data recovery issues due to external breaches<br>CSC 19: External incident response |

**Table 2.** Examples of cybersecurity activities in quality cost categories.

**Quality Costs and Organizational Maturity**

There has been little work relating the relative levels of quality costs in each category to organizational maturity, and no work to date relating cybersecurity costs of quality to the maturity of an organization's cybersecurity risk management. Several very limited papers exist that show patterns in quality costs for one organization, but these are of limited value because generalizability is low. Of the two models that do exist in the literature (Knox, 1993; Sower et al., 2007), both indicate nearly the same results: that as an organization matures, total quality costs decrease, but cost of prevention increases to make this possible.

This is illustrated in Figure 7 from the paper by Knox (1993), who explored quality costs in relation to the maturity of software development processes only. Both appraisal costs and prevention costs follow a similar pattern as well: they peak when an organization has systematic, repeatable processes and regular, effective training, but then decrease as feedback from learning and integration becomes stronger.

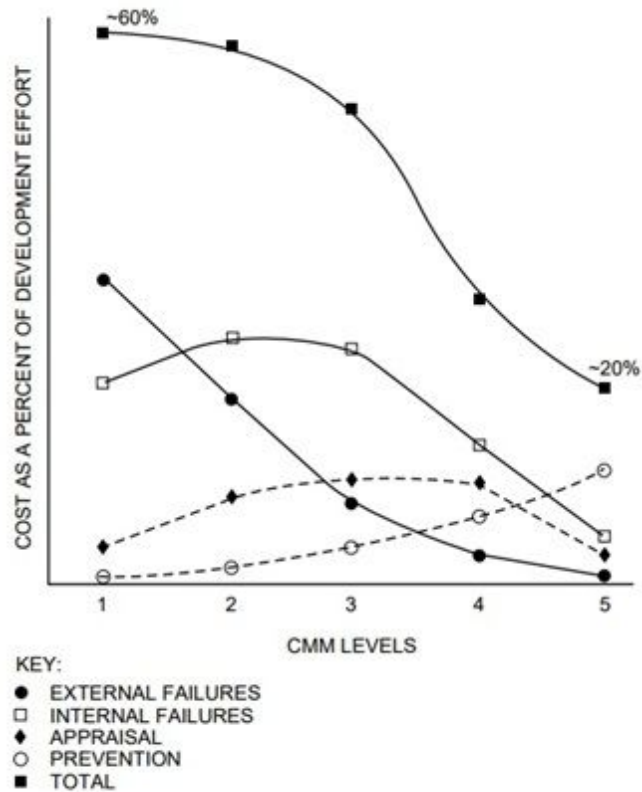

**Figure 7.** Quality costs classified by organizational maturity (CMM 1=low, CMM 5=high). From Knox (1993)

The value of cybersecurity cost of quality will increase as studies are performed to link measured costs from organizations in different industries to maturity levels. Most likely, this maturity would be assessed using tools like the three levels of the Cybersecurity Capability Maturity Model (C2M2) or the implementation tiers of the Baldrige Cybersecurity Excellence Builder (BCEB), both of which are designed to work well with the NIST CSF. Because the BCEB was not specifically designed to assess maturity, newly proposed structures may be required. (Almuhammadi & Alsaleh, 2017)

# Mapping of NIST CSF to Quality Costs

The primary contribution of this paper is a mapping of the 98 subcategories in the NIST Cybersecurity Framework (CSF) to the four main quality cost categories. By categorizing the elements of the NIST CSF at this level, organizations that use the CSF or a similar model can more easily adopt quality costs as a metric.

## Methodology

The mapping was produced using a consensus process with two observers (the authors). Independently, each observer classified the 98 NIST CSF subcategories into the four quality cost categories. Multiple classifications were allowed. To distinguish appraisal activities from prevention in the NIST CSF, keywords like "audit," "assess," and "verify" were used. To distinguish between appraisal and failures, the likelihood of the appraisal to be ordinary (that is, to *not* detect a failure) was considered. To distinguish between internal failures and external failures, the likelihood of information about a breach reaching a non-insider audience was considered.

Interrater reliability was assessed using Cohen's kappa, with the guideline outlined by Landis and Koch (1977), where the strength of the kappa coefficients =0.01-0.20 slight; 0.21-0.40 fair; 0.41-0.60 moderate; 0.61-0.80 substantial; 0.81-1.00 almost perfect. Because there were so many subcategories in the NIST CSF that map to both internal and external failures, a fifth category was added for the kappa calculation to reflect this combination. For elements in which one rater marked one quality cost category, and the other rater marked that category plus an additional category, one agreement mark was given for the category in agreement, and one disagreement mark to reflect the disagreement. As a result, 106 ratings were used to compute a kappa of 0.64, indicating substantial agreement. Of the five categories that were assessed for agreement, the strongest agreements were in prevention and the elements that combined internal failures and external failures. The least agreement was for classifications into the internal failures category.

Although a kappa of 0.64 suggests substantial agreement, the next step involved discussing classifications where there was disagreement to determine how a consensus recommendation could be achieved. There were 23 elements that required consensus determination; 7 were resolved easily, and 16 required more extensive discussion. The final mapping is shown in Table 3 with annotations. The analysis indicates that greater resolution is needed in the accounting system to accommodate for quality costs associated with several of the NIST CSF subcategories, including:

- **PR.IP-4:** Backups of information are conducted, maintained, and tested periodically: Conducting backups (Prevention) should be accounted for separately from testing backup processes (Appraisal) to see if they are functioning as anticipated.

- **PR.PT-1:** Audit/log records are determined, documented, implemented, and reviewed in accordance with policy: Auditing and verification (Appraisal) should be tracked separately from developing processes for logs (Prevention).
- **RS.RP-1, RS.CO-2 through RS.CO-5, RS.AN-3, RS.AM-4, RS.MI-1 and RS.MI-2** need to be accounted for by organizations in more detail, depending upon whether the work was associated with an internal or external failure.
- **RS.MI-3:** Newly identified vulnerabilities are mitigated or documented as accepted risks. This item is complex, encompasses many activities, and may even need to be revisited in the next update of the NIST CSF. When newly identified vulnerabilities are identified, that is an appraisal activity, but when steps are taken to prevent breaches before they occur, that is prevention. Furthermore, the identification of the vulnerability may take place in the context of responding to an internal failure or an external failure.

In addition, care should be taken when charging any activities to the subcategories in the Detect (DE) function. Detection occurs prior to the determination of a breach, and does not depend upon the audience that is impacted by that breach. Costs associated with the outcomes of these detections should be accounted for by subcategories in the Respond (RS) and Recover (RC) functions.

[INCLUDE SPREADSHEET on "FINAL MAPPING" tab AT https://docs.google.com/spreadsheets/d/1t6_1DLJHJqzA7zB7DM1GYEo7IMm_JVi4zfg-aQJbF7A/edit?usp=sharing AS TABLE 3]

**Table 3.** Mapping of the NIST CSF to the four quality cost categories.

**Application**

Using the mapping in Table 3, any work accounting system (such as an organizational Work Breakdown Structure or WBS) that is aligned with the NIST CSF can be adjusted to easily report cybersecurity cost of quality. The only change that may be required is to split some elements of the NIST CSF into two so that the appropriate quality cost categories can be assessed. Alternatively, if this is not feasible, an organization can create one category called "rework" in which all costs related to recovery from internal and external failures are grouped. Cost of rework may be just as useful a metric, particularly for organizations where a process approach and cybersecurity risk management are not as mature.

The organization can present this data as bar charts or Pareto charts, or can track the evolution of these values over time on segmented bar charts or in time series. Trends should be examined and discussed by staff and those with decision-making authority to identify opportunities for improvement. All values should be considered with respect to the maturity of the organization, in terms of quality management, risk management, and cybersecurity management. Most importantly, all organizations will require a period of adjustment and calibration when adopting a quality cost approach to help continually improve cybersecurity.

## Discussion & Conclusions

This article presented a mapping of the 98 subcategories in the Framework Core of the NIST Cybersecurity Framework to the four categories of quality costs (prevention, appraisal, internal failure, and external failure). The value of reporting costs of cybersecurity in terms of quality costs lies less in the levels themselves, and more in how the values relate to one another, change over time, and change in response to changes in strategy, organization, or cybersecurity investments. The primary limitation of this study is that the practical applicability of the model can only be assessed through future work on a broad scale: implementation at different organizations, case studies, and empirical research.

Based on this exercise, it became apparent that the NIST CSF does not distinguish between internal and external failures, and this is critical for managing the costs of cybersecurity - unless an organization chooses a simpler quality cost model, and groups internal and external failures to track cost of rework. Furthermore, there are a few subcategories in the NIST CSF that should be segmented in work accounting systems to reflect whether those tasks are being performed as preventive measures, or to test and appraise preventive measures. Organizations may also wish to customize the mapping to better align with their unique systems, structures, and work processes.

Using cybersecurity cost of quality, organizations can answer questions like:
- **Is there enough emphasis on prevention?** Typically, unless an organization's process maturity is high, this category will be associated with the greatest costs. If costs of prevention do not occupy the greatest proportion of quality costs, and
- **Is there too much or too little time being spent on appraisal (testing)?** If too little time is spent on appraisal activities, the cost of rework will be high. An excessively low cost of rework coupled with a high cost of conformance suggests that too much time is being spent on appraisal.
- **Is testing aggressive enough that it is triggering internal failures, or is the cost of internal failures small or nonexistent?** This is an indication that appraisal efforts should be made more rigorous.

Key questions to be answered in future research include:
- What are the theoretical and empirical relationships between the amount spent on quality cost categories (Prevention Appraisal, Internal Failure, External Failure) and the amounts spent on each function in the Framework Core (Identify, Protect, Detect, Respond, Recover)?
- Does the distribution of cybersecurity costs of quality change in a systematic way as an organization's cybersecurity risk management system matures?

- Does the distribution of cybersecurity costs of quality change in a systematic way as an organization's software lifecycle management matures?
- Do the costs of external failures increase proportionally as the attack surface expands?
- Are the costs of external failures similar for all organizations? Live benchmarking could help organizations identify whether they are being targeted for attacks.

Cost of quality is a time-tested and well documented model for identifying and assessing opportunities for improvement that will result in cost savings. By extending this model to cybersecurity risk management using a framework that is well known and widely applied, future empirical research that can provide cost benchmarks and additional methods for anomaly detection is enabled.